\documentclass[12pt,preprint]{aastex}
\shortauthors{Okamoto \& De Pontieu}
\shorttitle{Propagating waves along spicules}
\begin{document}

\title{Propagating waves along spicules}
\author{\textsc{
Takenori J. Okamoto,$^{a}$$^{b}$\footnote{JSPS Research Fellow and Visiting Scholar of Stanford University.}
Bart De Pontieu$^{b}$
}}
\affil{
$^{a}$National Astronomical Observatory, 2-21-1 Osawa, Mitaka, Tokyo 181-8588, Japan\\
$^{b}$Lockheed Martin Solar and Astrophysics Laboratory, B/252, 3251 Hanover St., Palo Alto, CA 94304, USA
}
\email{joten.okamoto@nao.ac.jp}

\begin{abstract}
Alfv\'enic waves are thought to play an important role in coronal heating and acceleration of solar wind. Here we investigated the statistical properties of Alfv\'enic waves along spicules (jets that protrude into the corona) in a polar coronal hole using high cadence observations of the Solar Optical Telescope (SOT) onboard \emph{Hinode}. We developed a technique for the automated detection of spicules and high-frequency waves. We detected 89 spicules, and found: (1) a mix of upward propagating, downward propagating, as well as standing waves (occurrence rates of 59\%, 21\%, and 20\%, respectively). (2) The phase speed gradually increases with height. (3) Upward waves dominant at lower altitudes, standing waves at higher altitudes. (4) Standing waves dominant in the early and late phases of each spicule, while upward waves were dominant in the middle phase. (5) In some spicules, we find waves propagating upward (from the bottom) and downward (from the top) to form a standing wave in the middle of the spicule. (6) The medians of the amplitude, period, and velocity amplitude were 55 km, 45 s, and 7.4 km~s$^{-1}$, respectively. 
We speculate that upward propagating waves are produced near the solar surface (below the spicule) and downward propagating waves are caused by reflection of (initially) upward propagating waves off the transition region at the spicule top. The mix of upward and downward propagating waves implies that exploiting these waves to perform seismology of the spicular environment requires careful analysis and may be problematic.
\end{abstract}


\section{Introduction}

Coronal heating and the acceleration of the solar wind are unsolved problems in solar physics. The propagation of Alfv\'en waves along magnetic field lines is one of the candidate mechanisms to carry energy to large distances from the surface and heat the coronal plasma. Many theoretical models have been developed for coronal heating based on waves that are excited in the photosphere or chromosphere \citep[e.g.,][]{hol82, tu97, axf99, suz05, cra07, ant10}. 

It has proven difficult to show from observations that waves are driving coronal heating. It is clear that three components should be found: the generation, propagation, and dissipation of waves. Low-frequency transverse waves propagating in the corona have been reported using line-of-sight Dopplergrams from ground-based coronagraphs (\citealt{sak02, tom07, tom09}) and spectrometers in orbit \citep[][and references therein]{ban10,mci11}. Here we focus on waves along chromospheric features embedded in the corona because chromospheric observations can be obtained at high enough spatial resolution to directly resolve the transverse motion of field lines as a result of Alfv\'en waves. While there have been earlier reports (from ground-based data) of the presence of transverse waves along spicules \citep[for a review, see][]{zaq09}, filtergraph observations with the \emph{Hinode} satellite \citep{hinode} for the first time clearly resolved small scale oscillations in prominences \citep{oka07} and spicules \citep{dep07}. These oscillations were identified as low-frequency Alfv\'enic waves in the coronal volume. In the latter cases, it is not clear whether these waves are actually propagating waves or standing waves, because the phase difference of the oscillations along the field lines were not resolved. More recently, \cite{he09} introduced 4 examples of spicules that had high-frequency upward propagating waves with very small amplitudes.

The purpose of this paper is to determine the properties of transverse waves for a large statistical sample of spicules. We obtained observations with high cadence, using the Solar Optical Telescope (SOT; \citealt{sot01, sot02, sot03, sot04}) onboard \emph{Hinode}, to resolve the phase difference of waves along spicules. We developed a technique for the automatic detection of spicules and investigated the observed waves. We present an analysis of the waves, and discuss their properties.

\section{Observation and Data Reduction}

The \emph{Hinode} satellite observed a coronal hole (CH) boundary at the south pole with a cadence of 1.6~s from 11:06 UT to 11:59 UT on January 29, 2011. We obtained 1995 images of off-limb spicules with the Ca \textsc{ii}~H-line filter (3968\AA, 3\AA\ bandwidth) of the SOT. The field of view is 55\arcsec$\times$55\arcsec (512$\times$512 pixel$^2$).

\begin{figure}
\epsscale{1}
 \plotone{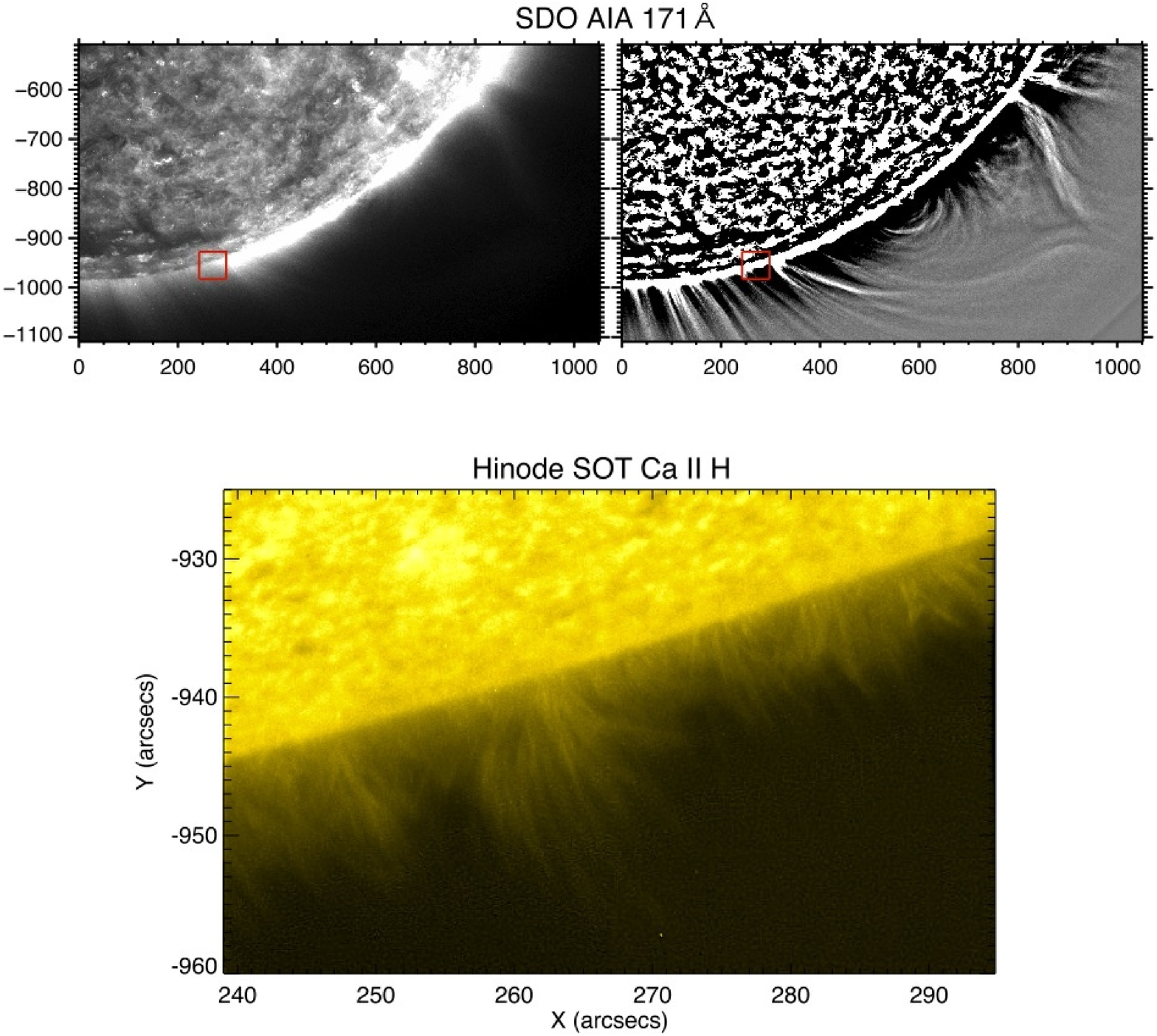}
    \caption{Context 171\AA\ image (left) from SDO/AIA at 11:00 UT on 29-Jan-2011 highlighting the location of the Hinode/SOT field of view (top panels, in red). The right panel shows the sum of 10 consecutive AIA images taken during 2 minutes which has been treated with unsharp masking to bring out the faint structures off-limb. Notice the large helmet streamer like configuration which has its roots to the north of the SOT field of view. This indicates that SOT was most likely pointed at a region that is part of the southern polar CH (i.e., open field conditions). The bottom panel shows a simultaneous Hinode/SOT Ca II H image illustrating the prevalence of chromospheric spicules.
}
    \label{fig1}
\end{figure}

Figure \ref{fig1} shows a Ca \textsc{ii}~H-line image at 11:06 UT and simultaneous coronal images taken with the 171\AA\ filter of the Atmospheric Imaging Assembly (AIA) onboard the \emph{Solar Dynamics Observatory}. While it is difficult to determine the boundaries of the CH because of line-of-sight confusion, the connectivity of the field lines visible off-limb (revealed by deep exposures and image processing of 171\AA\ images) indicates that the observed spicules are most likely located at the footpoints of open field lines.

To automatically detect a spicule and track it in time, we developed a technique based on the following steps:
(1) A radial density filter is applied to enhance visibility of faint features. 
(2) Each snapshot is averaged with 9 consecutive images to increase the signal-to-noise.
(3) A high-pass filter \citep[the \`{a} trous algorithm: ][]{waveletbook} is applied to emphasize bright features.
(4) The IDL procedure "thin.pro" is used to determine the coordinates of the medial axis of the bright features.
(5) To qualify as a spicule, each bright feature is required to be more than 4\arcsec\ (40 pixels) long and have a deviation from a straight line that is less than 5$^{\circ}$ over every 4\arcsec.
(6) The bright feature has to contain at least 30 pixels for which the horizontal displacement between consecutive snapshots (1.6 s apart) is within 1 pixel (at the same height). If there is no spicule detected in the timestep following the current timestep, the condition is applied to the medial axis of the next timestep that has a defined spicule (as long as it is within three timesteps of the current timestep).
(7) Each spicule is required to be longer than 8\arcsec\ at its maximum extent, and live longer than 40 s (25 timesteps).
(8) Each coordinate of the medial axes is calculated at the sub-pixel level by fitting a Gaussian. If the center axis deduced from the fitting is 4 pixels removed from the original medial axis, the Gaussian fit is not used. 

Using these procedures, we have obtained timeseries' of 96 different spicules. We visually inspected all spicules and rejected 7 that have irregular shapes, typically because multiple spicules are superposed along the line of sight. After removing these irregular spicules, we analyze the remaining 89 spicules (online animation A). While the detection method does not automatically classify the spicules as type I or II, most of the observed spicules are likely so-called type-II spicules \citep{dep07b}, because of their mostly upward motion, short lifetimes, high upward velocities, and occurrence in a CH.

\section{Results}

\begin{figure*}
\epsscale{1.0}
 \plotone{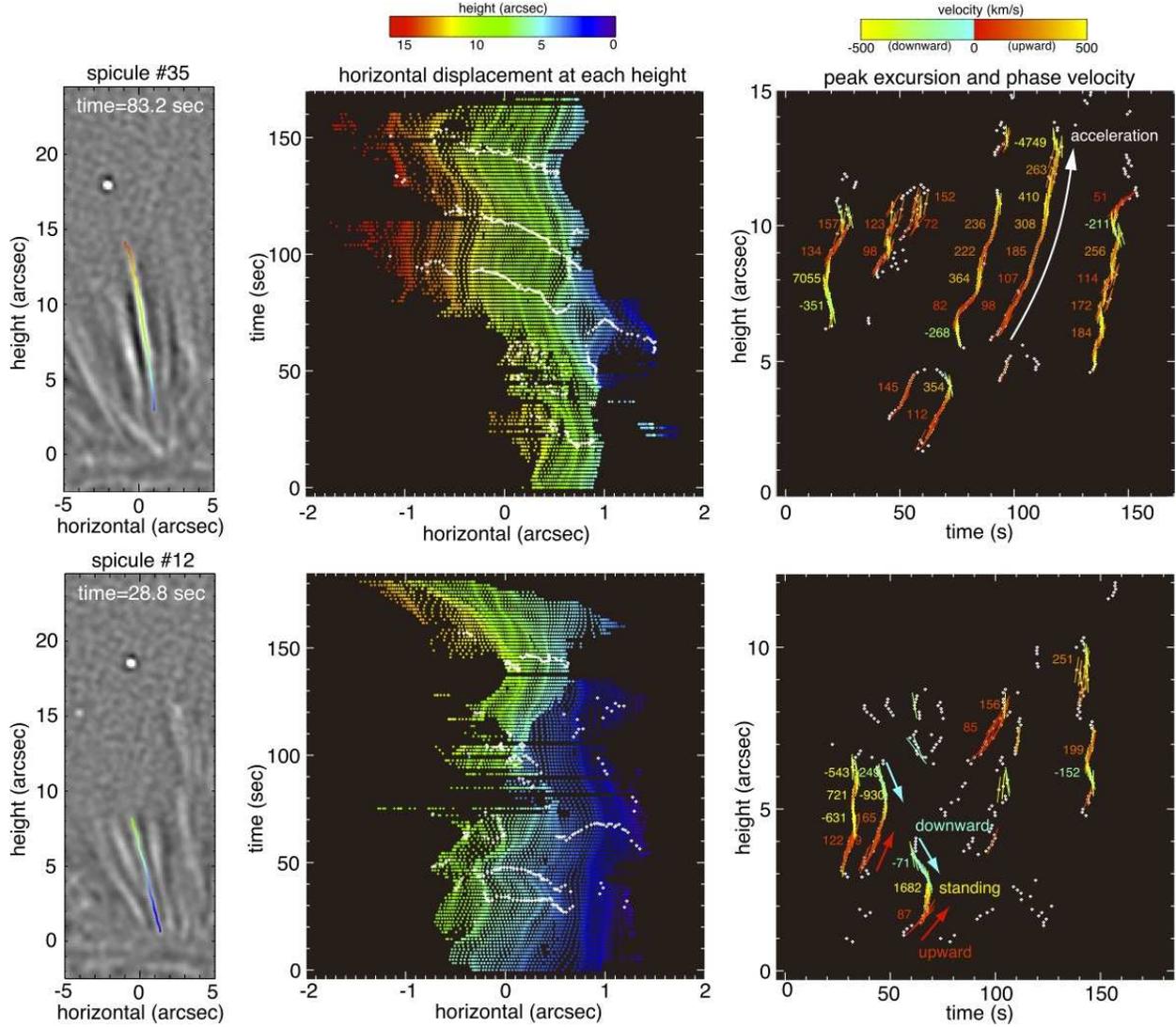}
    \caption{
Two examples of detected spicules and their transverse displacement.
Left: Snapshot. The detected spicule is highlighted with a colored line (color representing height).
Center: Time variation of horizontal displacements at each height of the spicule. The color represents height (same scale as left panel). Peak excursions of each wiggle are marked with white circular symbols.
Right: Peak excursions and the phase speeds in a time-height map. The numerals denote the values of the phase speed (km~s$^{-1}$). The upper panel shows that most of the waves propagate upward. One of the wave packets is seen to propagate with a speed that increases with height. The lower panel shows that there are a variety of wave packets along the spicule. In one case a wave packet propagates upwards from low heights and encounters a downward propagating wave packet (greater heights). A standing wave is produced in the middle as a result of superposition of the propagating waves.
}
    \label{fig2}
\end{figure*}

We study individual packets of transverse waves by analyzing the horizontal displacement of the medial axes of the spicules as a function of time and height (Figure \ref{fig2}). The middle panels show the horizontal displacements as a function of time. In both examples (top and bottom row), we can see oscillatory patterns as a function of time for each separate height (shown with color). To analyze the propagation, we focus on the peak excursions of the oscillations. The white circles in the middle panels indicate the location of peak excursion for an individual height. This peak excursion is derived separately for each height by fitting a Gaussian to the displacement vs. time curve for a timerange of 21 timesteps around the local peak in the horizontal displacement vs. time plot. We can easily see in the top-middle panel that several of the peak excursions move upwards (in height, i.e., color) with time (y-axis). This means there are upward propagating waves along this spicule. Here we define a set (i.e., for a range of heights) of peak excursions as one wave packet. We calculate the phase velocity of the wave packet at a certain height $h$ by performing a linear fit of the peak excursion curve (in the time vs. height domain) over 9 heights centered around height $h$. \footnote{To calculate the phase speed, we focus on individual excursions as opposed to performing a cross correlation on the whole displacement vs. time curve. The latter does not provide reliable values when a horizontal-time plot shows only one peak excursion.} 

The results are shown in the right panels of Figure \ref{fig2} with phase speed shown in colors and numerals. We can clearly see in the top-right panel that one of the wave packets propagates upward with a phase speed that increases with height (white arrow). In the bottom-right panels, we see an example of the interaction between upward (at low heights) and downward (at greater heights) propagating wave packets. The intermediate height range shows evidence of a wave packet with very high phase speed. We speculate that this is evidence for a standing wave, as a result of the superposition of the upward and downward propagating packets. While a pure standing wave would have an infinite phase speed, we cannot expect to measure such a speed in our data, given the uncertainties in the measurements. In addition, if the upward and downward propagating waves do not have exactly the same amplitude, we would observe a partially standing wave with very high (but not infinite) phase speed. Here, we define waves with speeds of more than 500 km~s$^{-1}$ as standing waves, because the chromospheric Alfv\'en speed is likely to be significantly less than that. The interaction of the waves is also illustrated in online animations B and C, with red and blue representing upward and downward propagating waves, respectively, and yellow for standing waves.
 
Next, we study the dependence of the phase speed with respect to height and time (since the beginning of the spicule). Figure \ref{fig3} shows the distribution of the phase speeds. On the horizontal axes we use the time lag between different heights, which is equal to (velocity)$^{-1}$, instead of the phase speed (shown on the top horizontal axis). The left panel indicates the height dependence of the phase speed, with the diamond symbols marking the median phase speed for each height range. The median for the phase speed along the whole length of the spicules is 270$\pm$30 km~s$^{-1}$. This is roughly consistent with estimates for the Alfv\'en speed in the quiet Sun, where the magnetic field strength is estimated to be 10 G \citep{tru05} and the plasma number density is of order $10^{10}$ cm$^{-3}$ \citep{bec68}. The median phase speed clearly increases with height and reaches values in excess of 1000 km~s$^{-1}$ at greater heights (i.e., standing waves dominate). The right panel indicates the dependence of the phase speed on the time elapsed since the spicule was first detected. We note that the start time is not the time of spicule formation, but the time at which the spicule fulfills our length criteria. We find a tendency for higher phase speeds to occur during the earlier phase, 
then upward waves are dominant, and finally the phase speed become higher again towards the end of the spicule's life.

We also measured the wave periods and amplitudes (Figure~\ref{fig4}) for wave packets that show at least two peak excursions at each height. The medians of amplitude and period are 0.077$\pm$0.072\arcsec\ (55 km, $\delta x$) and 45$\pm$30 s ($P$), respectively (with the standard deviation as error). Assuming a sinusoidal motion, we can calculate the median of the velocity amplitude ($v = 2 \pi \delta x / P$) as 7.4$\pm$3.7 km~s$^{-1}$. We can also estimate the velocity amplitude from a least squares fit of a linear function between the period and the amplitude. The best fit occurs for a velocity amplitude of 8.1 km~s$^{-1}$. This value is consistent with the result deduced from the assumption of a sinusoidal oscillation. Using these parameters, and assuming a plasma number density of $10^{10}$ cm$^{-3}$, we would roughly estimate the Poynting flux to be 2.5$\times 10^5$ erg cm$^{-2}$ s$^{-1}$ if the filling factor were one. It is however difficult to assess the filling factor of the waves using current data.

\section{Discussion}

We investigated the lateral motions of spicules observed with \emph{Hinode}/SOT, and detected numerous standing (20\%), and upward (59\%) and downward (21\%) propagating waves. 

The phase speed increases significantly with height along the spicules. This could be caused by a combination of two factors. First, at greater heights the Alfv\'en speed is higher since the plasma density is lower. If we assume that the plasma number density is roughly proportional to the observed intensity, the ratio of Alfv\'en speed at every 5\arcsec\ compared to that at the surface would be 1.00, 1.99, 2.95, and 3.39 (164, 327, 484, and 556 km s$^{-1}$ with 164 km s$^{-1}$ at 0\arcsec--5\arcsec). Considering that the magnetic field strength is expected to decrease with height, the range of phase speeds at heights below 15\arcsec\ thus seems reasonable. However, the phase speeds at heights above 15\arcsec\ seem too high to be explained by the change in Alfv\'en speed. A possible explanation is that the upward and downward propagating waves are more often superposed at greater heights. This would arise naturally in a scenario in which numerous waves are excited at lower heights, reflected at the transition region (TR) at the top of spicules, resulting in downward propagating waves at greater heights (just below the top). As a result, we would observe a combination of upward and downward propagating waves near the top of spicules. Given the high reflection coefficient of the TR \citep[see, e.g.,][]{hol82}, the downward propagating waves are expected to have amplitudes roughly equal to those of the upward propagating waves. However, some leakage of the wave energy does occur into the corona, so that the amplitudes are not quite equal. This means that superposition of the counterpropagating waves would lead to (partially) standing waves with very high phase speeds. This is exactly what we observe. We note that the spicules do not live very long (especially compared to the wave periods). This means that the reflected waves statistically will not reach back down to the bottom of the spicules. In summary, the combination of the increasing Alfv\'en speed and the reflection off the top of the spicule provide a good explanation for the observed change in phase speed with height. It is clear from our observations that the observed phase speed is not necessarily equal to the local Alfv\'en speed, and may in fact be very different from it because of the strong superposition of upward and downward propagating waves. Our results indicate that seismology of the spicular environment using Alfv\'en waves requires not only very careful analysis, but may in fact be highly problematic and unreliable.

This picture is confirmed when we look at the dependence of the phase speed on time: the phase speed decreases from the early phase to the middle phase, followed by an increase during the later phase of the spicule lifetime. While the start time is somewhat arbitrary, the definition is the same for all spicules, so it is reasonable to interpret the change as a function of time since spicule initiation. During the early phase, the spicules are shorter so that the TR is also located at lower heights. Hence, upward waves excited in the lower atmosphere would reach the TR sooner and are immediately reflected. This can only occur if the Alfv\'en speed ($\sim$ several hundred km~s$^{-1}$) is greater than the expansion velocity of the spicule ($\sim 50-100$ km~s$^{-1}$), which seems reasonable. Therefore, most wave signals at this stage consist of both upward and downward propagating waves, and the observed phase speed is higher. During the middle phase of the spicule lifetime, upward propagating waves are dominant. This could arise in a dynamic situation where wave packets are continuously propagating upward from the bottom, with the top of the spicule moving upward, and the phase speed is determined over the whole length of the spicule. During the middle phase there is a lower probability of the reflected waves reaching all the way back to the bottom of the spicule, so that there is a higher ratio of upward propagating waves. Towards the end, the reflected waves have had the time to propagate all the way back to the bottom (again resulting in a mix of upward and downward propagating waves). This scenario is speculative: the spicular environment is highly dynamic and the mix of waves will sensitively depend on the difference between the Alfv\'en speed and the velocity of the spicule (or the TR), and on the steepness of the TR. Our observations should be compared with and will help constrain detailed numerical models of Alfv\'enic disturbances on chromospheric spicules \citep[see, e.g.,][]{Mat10}. 

Such models will also have to explain why we see that upward propagating waves are three times more likely than downward propagating waves. At first glance, this may be seen as inconsistent with the idea that the TR has a high reflection coefficient \citep{dav91}. Perhaps the TR is not as sharp as traditionally assumed \citep[e.g., the hot plasma located beyond the height of cool spicules that was recently reported by][]{dep11}? Or perhaps the dynamic spicular environment with its competition between the upward spicule velocity and the Alfv\'en speed explains the dominancy of upward propagation?

\begin{figure*}
\epsscale{1.0}
 \plotone{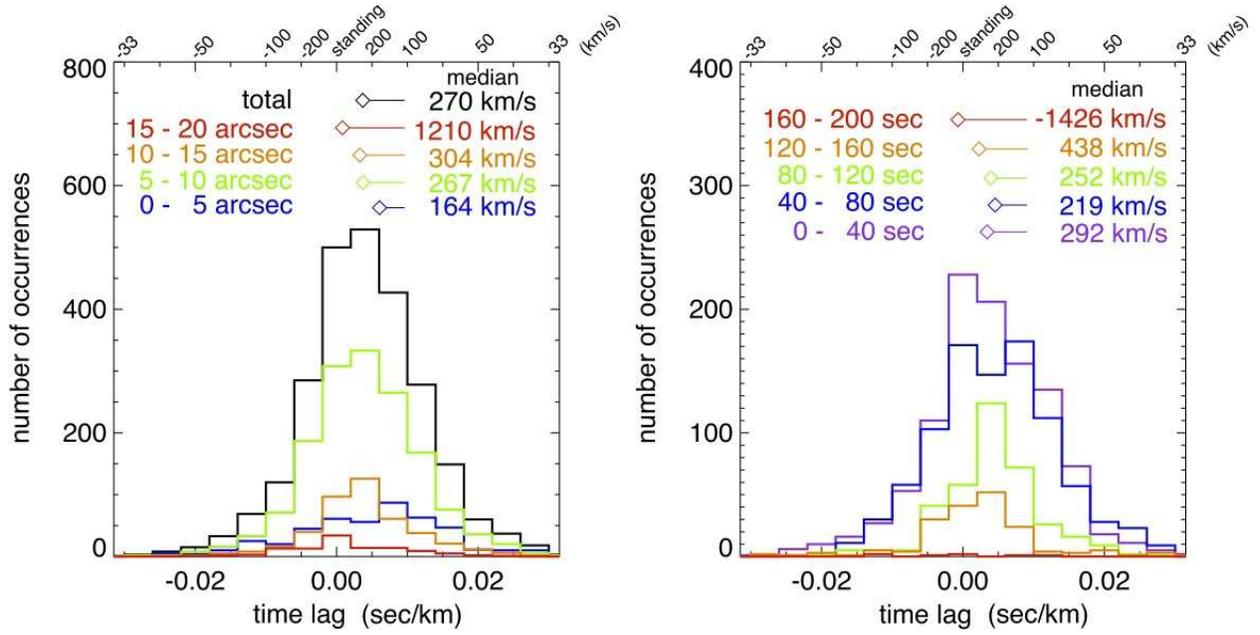}
    \caption{
Height and time dependence of the measured phase speed distributions.
Left: Histogram of phase speed binned for different height ranges (separated by five arcseconds) above the solar surface. The diamonds indicate the median values of the phase speed at each height range.
Right: Same as the left panel, but now for different time periods since the first detection of the spicules (binned every 40 seconds).
}
    \label{fig3}
\end{figure*}

\begin{figure}
\epsscale{1.0}
 \plotone{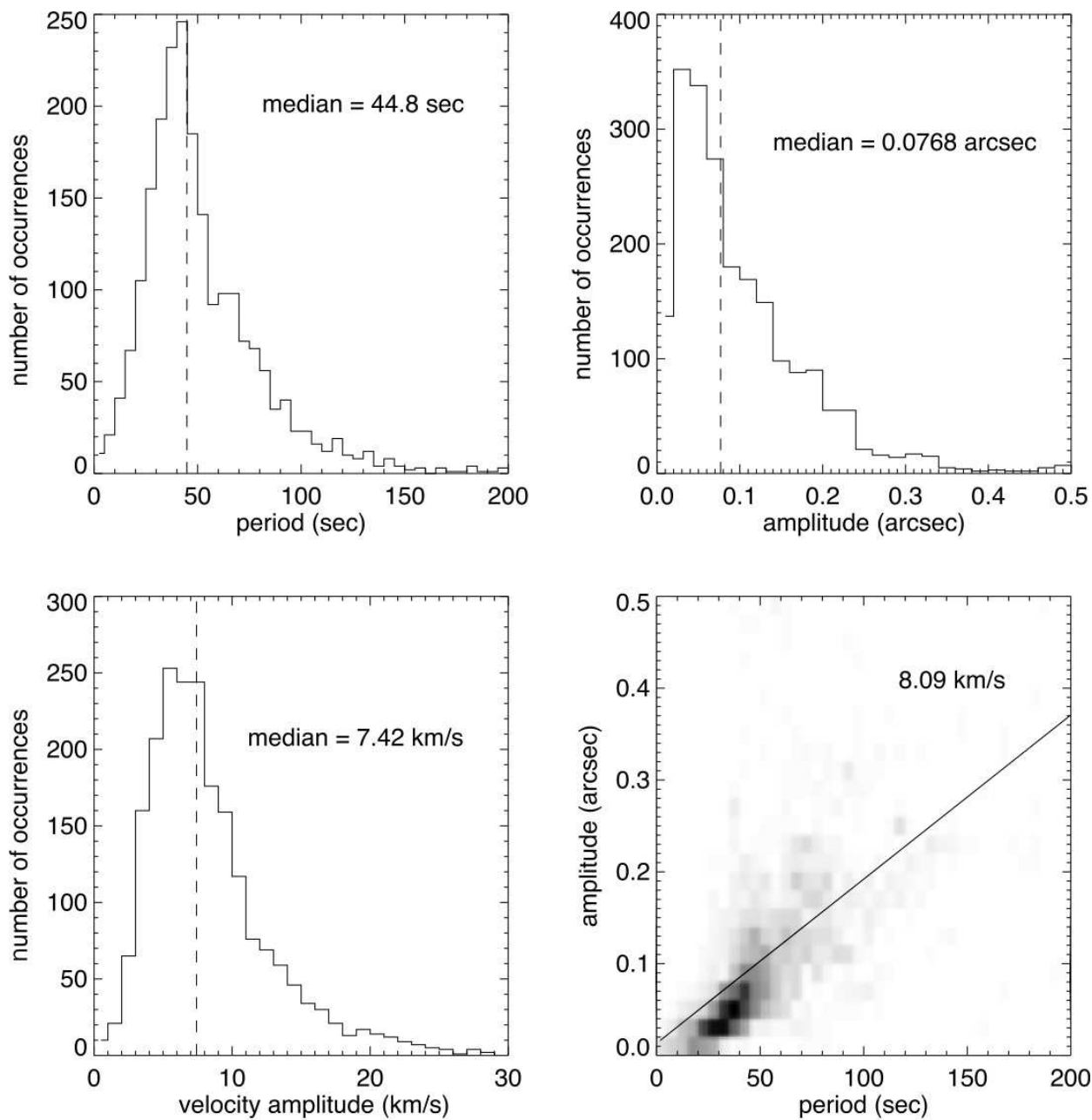}
    \caption{
Histogram of the measured amplitudes and periods (top row), and velocity amplitude (bottom left). Dashed lines show the median value. The velocity amplitude is determined from the amplitude and period by asssuming sinusoidal oscillations. The bottom-right panel shows a scatter plot of period and amplitude. The full line is a best fit which shows that a velocity amplitude of about 8 km~s$^{-1}$ fits best. This value is consistent with the median of the velocity amplitude deduced from the sinusoidal oscillation assumption (bottom left).
}
    \label{fig4}
\end{figure}

In our analysis, we have neglected the inclination of spicules. Since almost all of the detected spicules were within less than 10$^{\circ}$ from the vertical, the effect on the phase speed is small (of order 10\%). We also do not know where each spicule is rooted: some are located in front (or behind) of the limb, while others are located right at the limb. As a result each height measurement is actually a mix of heights above the solar surface, so that the phase speed distribution at lower heights includes information from greater heights. Given that the distribution at greater heights includes only those heights and shows greater phase speeds (Figure \ref{fig3}), the phase speed distribution at low heights would actually shift to the right (i.e., more upward propagation) if the information from greater heights were removed. In that case, the phase speeds would be better represented by dominance of upward propagating waves with phase speeds of 100--200 km~s$^{-1}$. This is close to the values extrapolated to lower heights from the measurements by \cite{mci11}.

The waves have rather short periods ($<100$ s) and small amplitudes ($<0.3\arcsec$). This is an artifact of our analysis. Because we require at least one full ``wiggle'' (i.e., a half period) during the spicule lifetime, we cannot measure the propagation or properties of the lower frequency waves ($P>100$ s) that carry much more energy \citep{dep07}. These lower frequency waves are not accessible with the method used, but are present in our dataset as well. 

The observed high frequency waves carry a significant energy flux of order 2.5$\times 10^5$ erg cm$^{-2}$ s$^{-1}$, in principle enough to play a role in heating the corona \citep{wit77}. However, our observations suggest that most of these waves likely do not reach the corona in sufficient quantity because of the reflection off the TR. We note that low frequency waves are even more efficiently reflected, but carry a much larger energy flux into the corona \citep{dep07}.

\acknowledgements 

\emph{Hinode} is a Japanese mission developed and launched by ISAS/JAXA, with NAOJ as domestic partner and NASA and STFC (UK) as international partners. It is operated by these agencies in co-operation with ESA and NSC (Norway). This work was supported by KAKENHI (21$\cdot$8014) and Grant-in-Aid for ``Institutional Program for Young Researcher Overseas Visits'' from the Japan Society for the Promotion of Science. B.D.P. was supported by NASA grants NNX08AL22G and NNX08BA99G.




\end{document}